\documentclass[conference]{IEEEtran}
\IEEEoverridecommandlockouts
\usepackage{cite}
\usepackage{amsmath,amssymb,amsfonts}
\usepackage{algorithmic}
\usepackage{graphicx}
\usepackage{textcomp}
\usepackage{xcolor}
\def\BibTeX{{\rm B\kern-.05em{\sc i\kern-.025em b}\kern-.08em
    T\kern-.1667em\lower.7ex\hbox{E}\kern-.125emX}}
\usepackage{bm}
\usepackage{pgfplots}
    \usetikzlibrary{
        arrows.meta,
        spy,
    }
\usepackage{filecontents}
\usepackage[scientific-notation=true]{siunitx}
\usepackage[T1]{fontenc}
\usepackage{graphicx,array,amsthm,stfloats,upref}
\usepackage{color,soul}
\graphicspath{{figures/}}

\usepackage{cite}
\usepackage{mathtools}

\usepgfplotslibrary{polar}

\pgfplotsset{compat=1.13}

\begin{document}

\title{Determination of the Phase Centers \\ of Multi-Mode Antennas
}

\author{\IEEEauthorblockN{Sami Alkubti Almasri and Peter~A.~Hoeher}
\IEEEauthorblockA{\textit{Institute of Electrical Engineering and Information Technology} \\
\textit{Kiel University}\\
Kiel, Germany  \\
\{saaa, ph\}@tf.uni-kiel.de}
}

\maketitle

\begin{abstract}
In this paper, the displacement of the phase centers of multi-mode antennas is investigated. 
A mathematical approach is applied in order to calculate the positions of the phase center for each mode. 
Using electric far field samples obtained from electromagnetic field simulation data, the coordinates of the shifted positions of the phase center relative to the physical center of the antenna are determined for each mode.  
Numerical results show the dependency of the phase centers of an antenna prototype as a function of the considered angle of observation.
\end{abstract}


\section{Introduction}
Determining the phase center (PC) position in an antenna system and the effect of its displacement plays a significant role in system performance for a wide range of applications.
For instance, in case of parabolic reflectors, ideally the PC of the feed antenna should coincide with the focal point of the reflector \cite{Balanis07}.  
Other applications, where knowledge of the PC is crucial, are high accuracy ranging and positioning systems including radar and global navigation satellite systems.
The error introduced through an uncompensated displacement of the PC leads to a significant performance degradation of ranging systems \cite{Spilker96}.
Accordingly, a precise positioning system should not neglect the impact of phase center displacements.
The research conducted for locating the PC includes methods that either measure the PC location depending on experimental settings \cite{Zeimetz08} or calculate the PC location based on measured or simulated field data \cite{Kunysz10}.
Some of these methods are only applicable on a specific type of antennas (e.g horn antennas \cite{Muehldorf70} and logarithmic-periodic dipole arrays \cite{Rong07}).
While others propose algorithms that are applicable universally regardless of the type of antenna in a 2-D plane \cite{Betjes07}, or in 3-D space \cite{Haerke17,Basta09}.
The latter studies not only take the antenna itself into account but the whole radiating system as a unit, i.e. including the reflecting and scattering effects resulting from objects surrounding the plain antenna body.

Multi-mode (MM) antennas are characterized by the fact that multiple modes can be excited simultaneously and independently given a single physical radiator.
As opposed to classical antennas, MM antennas have $M$ PCs, exactly one per mode. 
The capability of employing planar MM antennas for positioning purposes has been studied in \cite{Almasri17, Poehlmann17WPNC, Poehlmann17CAMSAP, Poehlmann19, Almasri19}. 
Consequently, it is an interesting aspect to investigate the PC locations of MM antennas and inspect the displacement of each PC depending on the observation angle.

Original contributions of this paper are as follows: (i) we employ the method proposed in \cite{Basta09} to MM antennas, (ii) we calculate the PCs of a given MM antenna prototype, and (iii) we point out that the PCs of different modes are dislocated for a given observation angle in the 3-D space. 

This paper is organized as follows: Section~\ref{sec2} introduces the definition of the PC and points out the importance of the knowledge of its location for positioning applications. 
Section~\ref{sec3} briefly presents MM antennas basics. 
Section~\ref{sec4} brushes up on the PC determination method.
The application of the PC determination method is discussed in Section~\ref{sec5}.
Simulation results are shown in Section~\ref{sec6}.
Finally, conclusions are drawn in Section~\ref{sec7}.

\section{Antenna Phase Center}\label{sec2}
\begin{figure}
	  \centering
 	  \includegraphics[width=\columnwidth,keepaspectratio]{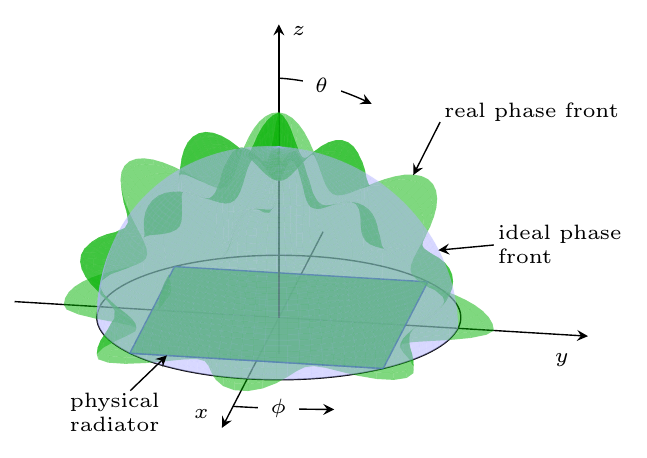}
   \caption{Phase fronts resulting from an ideal and a real PC.}
   \label{fig1}
\end{figure}

The PC of an antenna is defined as the point from which electromagnetic waves seem to be originating.   
A simple example is the theoretical isotropic antenna. 
In this case the phase center coincides with the physical radiator (point source), hence the electromagnetic waves appear to be radiating spherically outwards, generating equiphase fronts with the isotropic radiator being in the center of the progressing spheres. 
It is clear that it is far from logical to consider practical antennas to be point sources. 
Therefore, for realistic antennas it is not possible to find a single unique phase center that fulfills the phase consistency of the radiation surface in all directions \cite{Balanis07}.
Consequently, for realistic antennas the PC is spatially distributed on the structure of the antenna or even around it depending on the observation angle.
Furthermore, its position relies on multiple parameters such as frequency, polarization, antenna dimensions and structure.
However, according to IEEE standards \cite{ieeestandard}, a radiation point qualifies also as a PC if the consistency of the phase is fulfilled only over a portion of the radiation surface, at least where the radiation is significant. 
Thus, it is possible to find a PC for certain angular portions of the space so that the phase of the radiated waves is roughly constant within this specific range.
Fig.~\ref{fig1} visually explains the definition of the PC.
A planar physical radiator is assumed to be placed on the $xy$-plane. 
In the case of an ideal unique PC located in the physical center of the radiator, i.e. at the origin of the Cartesian 3-D space, the phase fronts would form a spherical surface in the far field. 
As opposed to that, the real PC is spatially distributed and it depends on the observation angle. 
As a consequence, the phase front is not spherical. 
Please notice that the phase pattern plotted in Fig.~\ref{fig1} was chosen arbitrarily and serves only for illustration purposes.
Among PC determination methods are a lot of practical methods and some theoretical ones.
For certain applications it is important to find the phase center position in order to consider it as a reference point of the antenna rather than the common convention of considering the physical center of the antenna structure.
It can therefore be seen as the reference point in an antenna system to which a distance measurement refers.

\section{Multi-Mode Antenna Basics}\label{sec3}
An MM antenna physically consists of a single antenna radiator that emanates multiple modes depending on the excitation applied on it. 
The concept behind the capability of MM antennas is provided by the theory of characteristic modes (TCM) \cite{Garbacz71, Harrington71, Chen15}.  
The principle of TCM can be described as the decomposition of the surface currents of an electric conductor into its orthogonal components. Each of these components excites a certain mode.
In other words, an MM antenna is able to emit multiple orthogonal radiation patterns simultaneously \cite{martens14}.
A powerful feature that offers a compelling diversity gain with a suitable designed MM antenna for MIMO applications \cite{Svantesson02}.
Furthermore, it has been proven that different modes exhibit less correlation compared to a linear array of the same size as a planar MM antenna \cite{Hoeher17}.
The latter feature of MM antennas motivated the investigation of their capabilities when applied for positioning purposes. 
The work published in \cite{Almasri17, Poehlmann17WPNC, Poehlmann17CAMSAP, Poehlmann19, Almasri19} demonstrates the suitability of MM antennas for positioning applications using different positioning methods as well as the tools necessary for applying conventional array signal processing techniques on MM antenna.
Nevertheless, the position of the phase centers of an MM antenna has not yet been investigated. 
It has been suggested in \cite{Almasri17} that the effective phase centers of the modes of a planar MM antenna are not spatially distributed for a given direction-of-arrival (DoA). 
However, the results of the following analysis presented here show that given a certain DoA, a significant displacement of the phase centers of different modes exists, and should ideally be compensated for accurate positioning applications. 

\section{Simplex-Based Phase Center Determination Method}\label{sec4}
The PC determination algorithm chosen for our analysis follows the method presented in \cite{Basta09}.
The phase of the electric field $\psi(\phi,\theta)$ of an antenna at a distance $R$ and angle $(\phi,\theta)$ in the far-field consists of three contributions:
\begin{equation}\label{eqtotalphase}
\psi(\phi,\theta) = \psi_\mathrm{i} + \psi_\mathrm{R} + \psi_\mathrm{s}(\phi,\theta),
\end{equation}
where $\psi_\mathrm{i}$ is the initial phase contribution and $\psi_\mathrm{R}$ is the shift caused by the propagation of the wave at distance $R$.
These two contributions are obviously constant. 
Hence, from now on they are treated as one constant phase term $\psi_\mathrm{c}$.  
The last contribution $\psi_\mathrm{s}(\phi,\theta)$ is the phase shift resulting from the displacement of the PC.
Assuming that the PC is displaced by the radial vector $\bm{r}$, and that the wave number vector in the direction of propagation is $\bm{k}$, then the PC phase shift can be expressed as
\begin{flalign}\label{eqphaseshift}
 \psi_\mathrm{s}(\phi,\theta) & =  \bm{k} \cdot \bm{r} && \\\nonumber
			& = k(\bm{e_x} +\bm{e_y} + \bm{e_z}) \cdot (x_\mathrm{PC}(\phi,\theta)\cos\phi\sin\theta \cdot \bm{e_x}+ && \\
			& \phantom{=~} y_\mathrm{PC}(\phi,\theta)\sin\phi\sin\theta \cdot \bm{e_y}+z_\mathrm{PC}(\phi,\theta)\sin\theta \cdot \bm{e_z}) && \\\nonumber
			& = k (x_\mathrm{PC}(\phi,\theta)\cos\phi\sin\theta+ &&\\\label{eqphasedetail}
			& \phantom{=~} y_\mathrm{PC}(\phi,\theta)\sin\phi\sin\theta+z_\mathrm{PC}(\phi,\theta)\sin\theta), 
\end{flalign}
where $k=2\pi/\lambda$ and $[x_\mathrm{PC},y_\mathrm{PC},z_\mathrm{PC}]$ are the wave number and the Cartesian coordinates of the PC displacement, respectively.
Notice that the PC coordinates are dependent as mentioned before on the observation angle.
When these coordinates become independent of the observation angle in the whole 3-D space, then the PC is placed in a single location and is unique for all direction, i.e. the phase fronts in the far field build a sphere with the PC at the center of it.
As mentioned in Section~\ref{sec2}, a unique PC does not exist for realistic antennas. 
Therefore, finding a unique PC only for a certain angular region of the 3-D space is normally sought. 
Considering the previous discussion and inserting (\ref{eqphasedetail}) into (\ref{eqtotalphase}), the total phases of the angular region $p$ can be expressed as  
\begin{flalign}\label{phasecompact}
\bm{\psi}(\bm{\phi}_p,\bm{\theta}_p) & = \bm{\psi}_\mathrm{c} + \bm{\psi}_\mathrm{s}(\bm{\phi}_p,\bm{\theta}_p) && \\
&= \bm{\psi}_\mathrm{c} + k (x_\mathrm{PC}\cos\bm{\phi}_p\sin\bm{\theta}_p + &&\\\nonumber
 & \phantom{=~} y_\mathrm{PC}\sin\bm{\phi}_p\sin\bm{\theta}_p+z_\mathrm{PC}\sin\bm{\theta}_p),
\end{flalign}
given that the $p$-th angular region $\bm{\phi}_p = [\phi_1,\phi_2,\dots,\phi_{N_p}]$ and $\bm{\theta}_p = [\theta_1,\theta_2,\dots,\theta_{N_p}]$ contains $N_p$ samples.
The objective is to find the PC coordinates $[x_\mathrm{PC},y_\mathrm{PC},z_\mathrm{PC}]$ subject to the constraint
\begin{equation}\label{constr}
\bm{\psi}_{\mathrm{eq}}(\bm{\phi}_p,\bm{\theta}_p) = \bm{\psi}(\bm{\phi}_p,\bm{\theta}_p) - \bm{\psi}_\mathrm{s}(\bm{\phi}_p,\bm{\theta}_p) \approx \mathrm{const},
\end{equation}
i.e., to find the PC coordinates that would optimally neutralize the phase shift to end up with an equiphase surface over the angular region $p$.
For this purpose the variance 
\begin{equation}\label{var}
 \sigma^2 = \sum_{n=1}^{N_p}(\bm{\psi}_{\mathrm{eq}}(\phi_n,\theta_n)-\mu_{\psi_{\mathrm{eq}}})^2 
\end{equation}
of the phase pattern $\bm{\psi}_{\mathrm{eq}}$ over the regarded angular region $p$ is minimized, where $\mu_{\psi_p}$ is the mean of the phase samples in the region $p$.
This nonlinear optimization problem is solved using the Nelder-Mead simplex algorithm.
It employs an iterative direct search approach towards convergence. 
For detailed information about the method the reader is referred to \cite{Lagarias98}.
 
\section{Application to a Multi-Mode Antenna Prototype}\label{sec5}
\begin{figure}
	  \centering
 	  \includegraphics[width=\columnwidth,keepaspectratio]{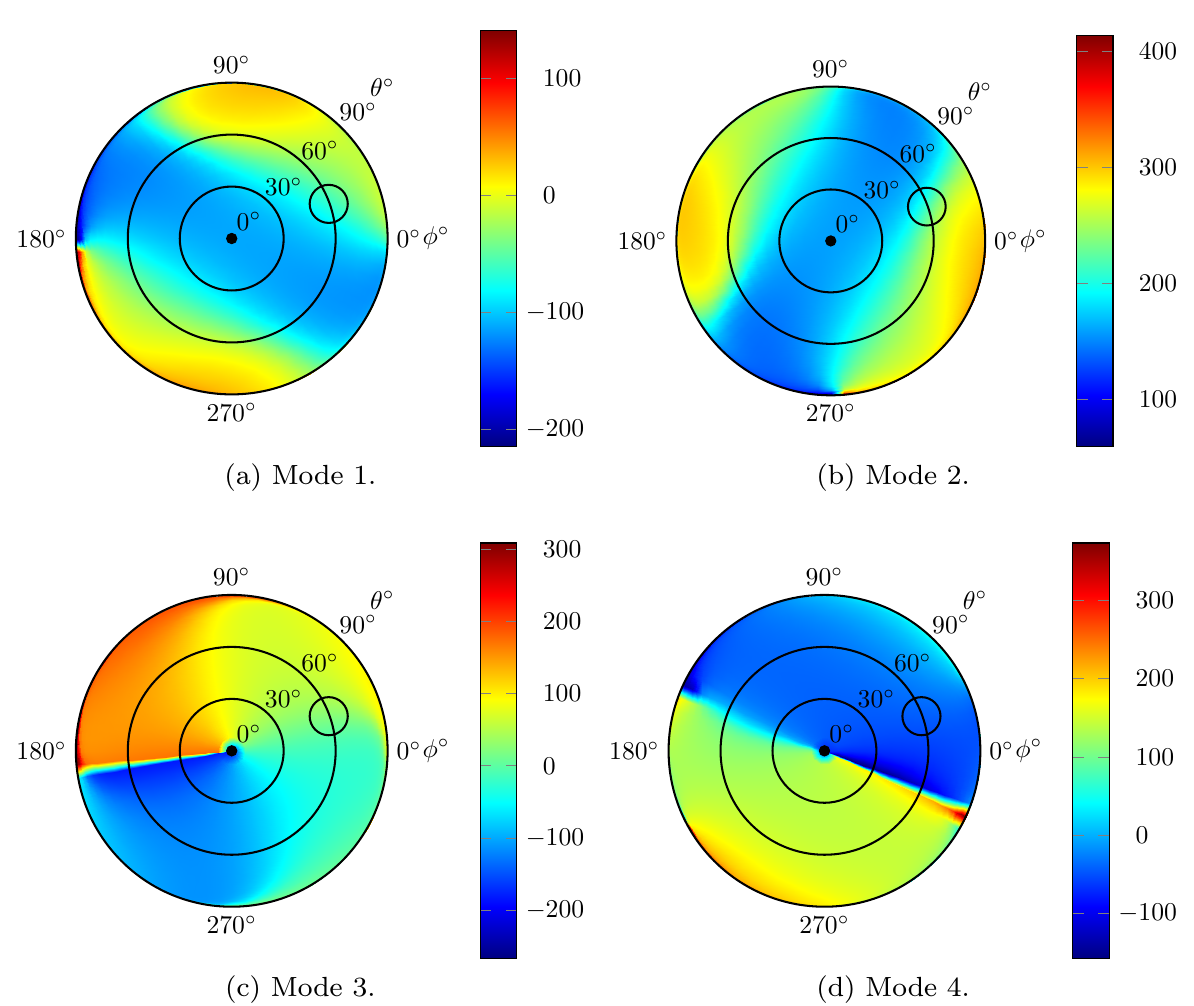}
    \caption{Phase patterns of the considered electric field component of the investigated MM antenna. The studied region $[\phi_0,\theta_0] = [20^\circ,60^\circ]$ is marked by a circle on each mode.}
    \label{fig2}
  \end{figure}

For the analysis of the phase center positions of an MM antenna, the planar prototype presented in \cite{Manteuffel16} is considered next in order to provide numerical results. 
The mentioned MM antenna is provided with 4 ports for exciting $M=4$ modes at a center frequency of $f_c=7.25$ GHz, i.e. $\lambda=0.0414$ m. 
The radiator dimensions are $30$ mm $\times$ $30$ mm ($ = 0.725\lambda \times 0.725\lambda$). 
For further technical details on the MM antenna design under investigation see \cite{Manteuffel16}.
Fig.~{\ref{fig1}} illustrates the considered coordinate system, the position of the antenna, and the definition of angles $\phi$ and $\theta$.
The MM antenna is placed on the $xy$-plane where the center of the planar radiator coincides with the center of the Cartesian coordinate system. 
The radiation takes place in the upper hemisphere, therefore the considered azimuth and co-elevation angles are defined as follows: $\phi\in[0^\circ,360^\circ]$ and $\theta\in[0^\circ,90^\circ]$.
For the following analysis, the right-hand circularly polarization (RHCP) component is assumed to be the dominant component and therefore it is the studied component in our analysis. 
As aforementioned, each component of the electric field fulfills the definition of the phase center, i.e., the method applied to the RHCP component can be successfully applied on any other polarization component.  
Fig.~{\ref{fig2}} shows the 3-D phase pattern of each of the four modes of the studied MM antenna.

\section{Simulation Results}\label{sec6}
\begin{figure}
	  \centering
 	  \includegraphics[width=\columnwidth,keepaspectratio]{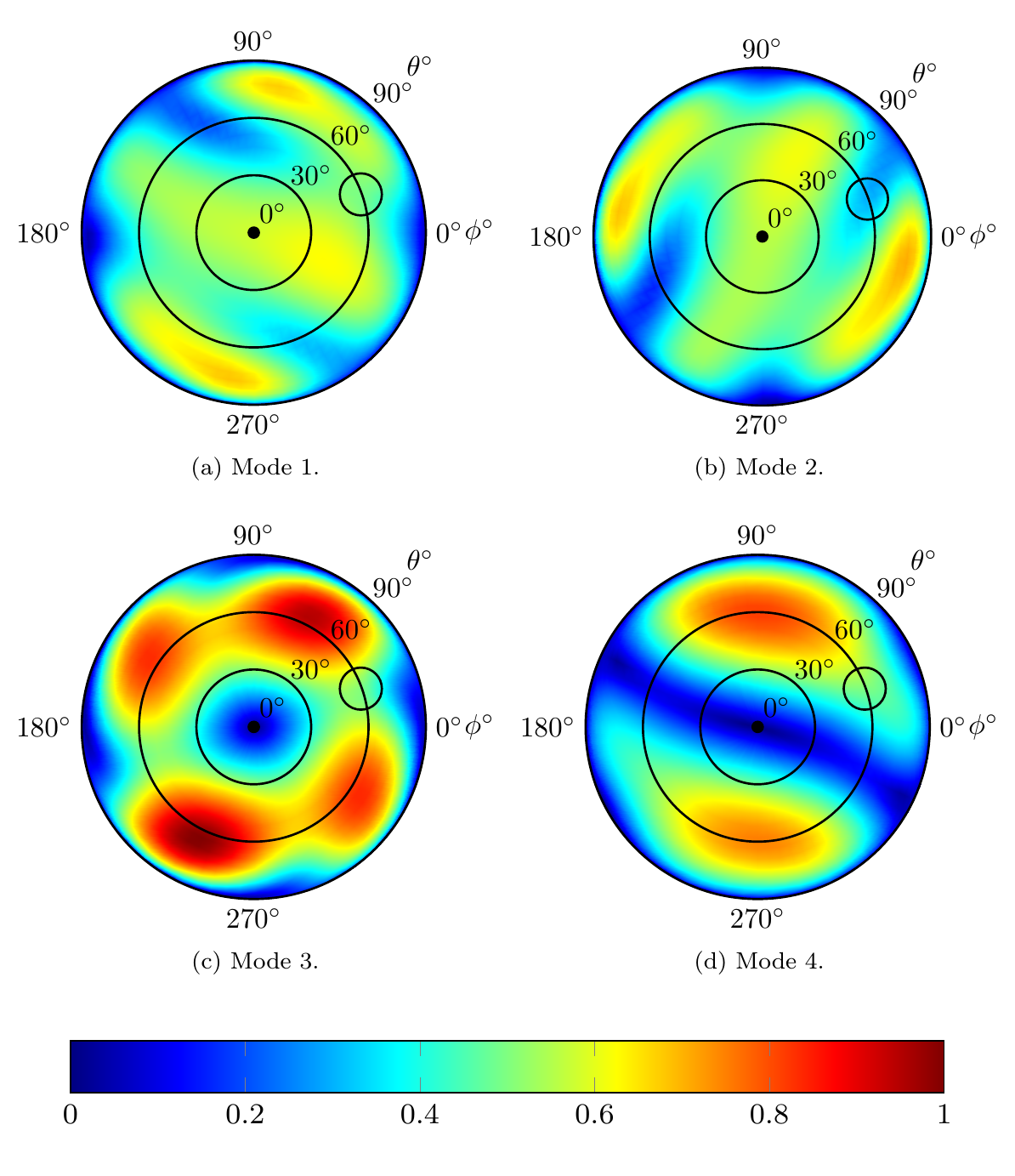}
    \caption{Normalized magnitudes of the considered electric field component of the investigated MM antenna. The studied region $[\phi_0,\theta_0] = [20^\circ,60^\circ]$ is marked by a circle on each mode.}
    \label{fig3}
  \end{figure}

In the following simulations, the PC of each of the four modes has been calculated separately. 
The used phase pattern samples have a resolution of $1^\circ$ for both $\phi$ and $\theta$ angles.
First, the phase samples are unwrapped along the $\theta$ direction to avoid large phase jumps, specially in the region considered for analysis \cite{Basta09}.
Fig.~\ref{fig2} shows the unwrapped phase patterns for each mode of the investigated MM antenna prototype.
Notice that the phase values go beyond the range $[-180^\circ,180^\circ]$ because of the unwrapping process.
Next the unwrapped phase samples within the area of interest are gathered and entered in~(\ref{var}) and the variance is minimized according to the Nelder-Mead simplex algorithm.
To illustrate the results, an angle of $[\phi_0,\theta_0] = [20^\circ,60^\circ]$ was taken.
An angular region with a diameter of $20^\circ$ around $[\phi_0,\theta_0]$ was considered. 
The mentioned angular region is depicted in Fig.~\ref{fig2} by a small circle.
It is recommended in \cite{Basta09} that the sampling area of the phase pattern exhibits approximately a consistent magnitude pattern for reasonable results.
As can be seen in Fig.~\ref{fig3}, the chosen sampling area fulfills this condition.
Table~\ref{table1} shows the displacement of the PC coordinates for the observation angle $[\phi,\theta]$.

\begin{figure}
	  \centering
 	  \includegraphics[width=\columnwidth,keepaspectratio]{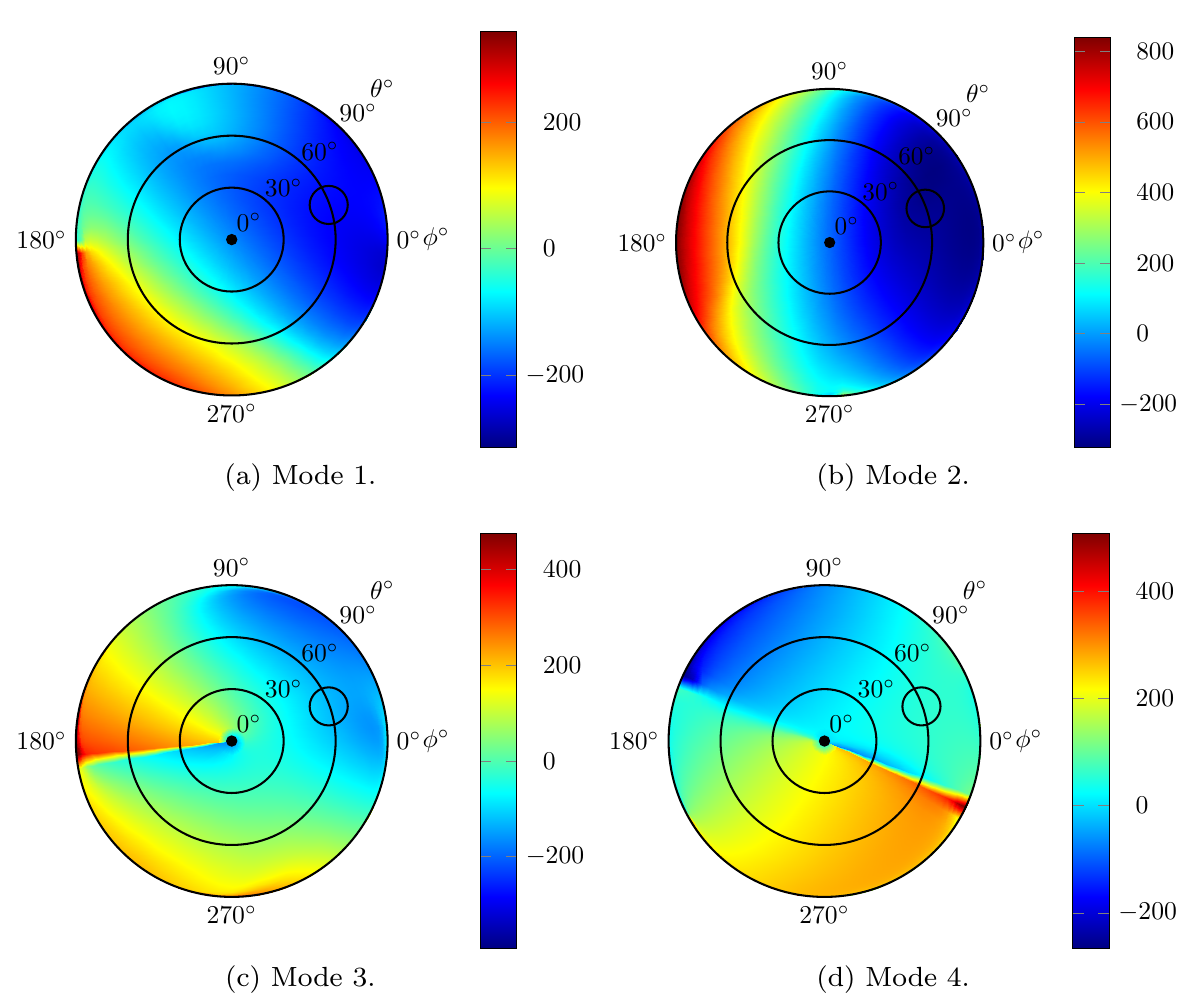}
    \caption{Phase patterns of investigated MM antenna after subtracting the phase shift caused by the PC displacement in the region around $[\phi_0,\theta_0] = [20^\circ,60^\circ]$.}
    \label{fig4}
  \end{figure}

\begin{figure}
	  \centering
 	  \includegraphics[width=\columnwidth,height=10cm]{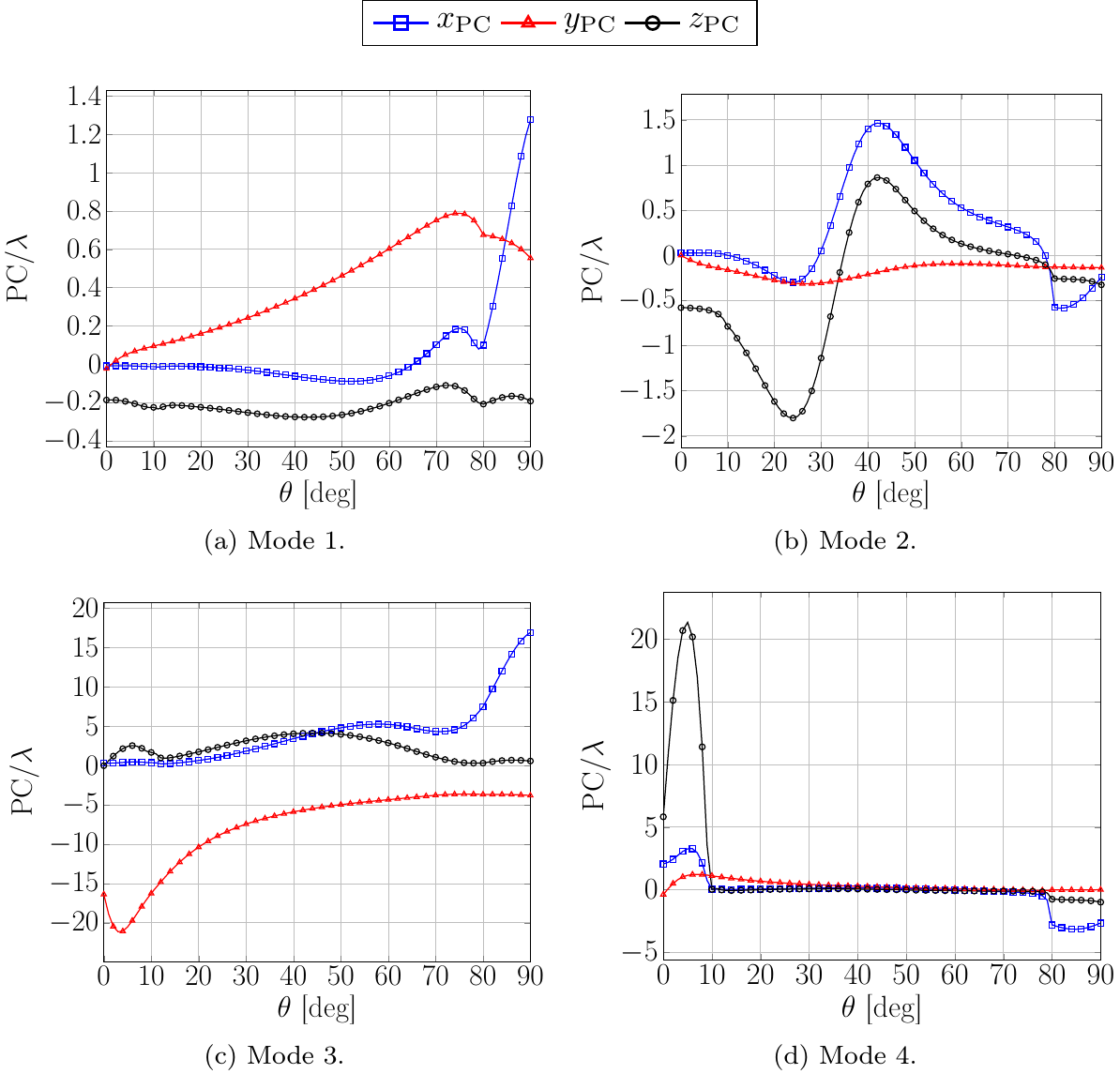}
    \caption{Coordinates of the phase centers as a function of $\theta$ in $\phi=0^\circ$ plane.}
    \label{fig5}
  \end{figure}

After calculating the phase center, one could use the resulting coordinates to calculate the phase shift $\bm{\psi}_\mathrm{s}(\bm{\phi}_p,\bm{\theta}_p)$ and insert it in (\ref{constr}) to calculate $\bm{\psi}_{\mathrm{eq}}(\bm{\phi}_p,\bm{\theta}_p)$.
Fig.~\ref{fig4} illustrates $\bm{\psi}_{\mathrm{eq}}(\bm{\phi}_p,\bm{\theta}_p)$ where it is clear to see that the phase pattern is approximately constant in the studied range.
\begin{center}
\begin{table}[h]
\caption{Calculated PC coordinates for observation angle $[\phi,\theta] = [20^\circ,60^\circ]$.}
    \begin{tabular}{ | l | l | l | l |}
    \hline
     & $x_{\mathrm{PC}}$ & $y_{\mathrm{PC}}$ & $z_{\mathrm{PC}}$ \\
     \hline
    mode $1$ & 20.508 mm & 16.080 mm &  3.417 mm\\
    \hline
    mode $2$ & 65.246 mm & 0.753 mm &  25.280 mm\\
    \hline
    mode $3$ & 17.596 mm & 30.482 mm &  -1.036 mm\\
    \hline
    mode $4$ & -11.735 mm & 9.373 mm &  -7.891 mm\\
    \hline
    \end{tabular}\label{table1}
    \end{table}
\end{center}
In addition, the behavior of the PCs of the four modes is depicted in Fig.~\ref{fig5} for the $\phi=0^\circ$ and $\theta=[0^\circ,90^\circ]$ planes.
The displacement of the PC coordinates has been illustrated relative to the wavelength.
The dependency of the PC coordinates of each mode on the observation angle becomes obvious from Fig.~\ref{fig5}. 
It is also noticeable that the PC of each mode is independent from the PCs of other modes.
These results proof that the displacement of the PCs of different modes is significant, therefore it should not be neglected when employing the MM antenna under investigation in precise positioning systems.
\section{Conclusion}\label{sec7}
Knowledge of the location and behavior of the phase center of an antenna system is important for high accuracy positioning systems.
In this paper, the PC locations have been calculated for an MM antenna prototype.
For that purpose, the PC coordinates that minimize the variance are determined over a certain angular region of interest.
The minimization is performed using the Nelder-Mead simplex algorithm. 
Numerical results show the dependency of the PC of each mode on the observation angle.
Furthermore, it has been proven that the PCs of different modes are independent of each other. 
The PCs are spatially distributed around the physical radiator as a function of the observation angle.
With this insight it is possible to compensate the error introduced by the PC displacement when applying an MM antenna for distance measurements, depending on the modes used in the transmission process.
\section*{Acknowledgment}
This work has been funded by the German Research Foundation (DFG) under contract number HO~2226/17-2. 
The authors would like to thank Prof.~D.~Manteuffel and Nikolai Peitzmeier from the University of Hannover, Germany, as well as Robert P\"ohlmann and Dr. Armin Dammann from DLR Oberpfaffenhofen, Germany, for collaborations on MM antennas.

\newpage

\bibliographystyle{IEEEtran}
\bibliography{references}

\begin{thebibliography}{10}
\providecommand{\url}[1]{#1}
\csname url@samestyle\endcsname
\providecommand{\newblock}{\relax}
\providecommand{\bibinfo}[2]{#2}
\providecommand{\BIBentrySTDinterwordspacing}{\spaceskip=0pt\relax}
\providecommand{\BIBentryALTinterwordstretchfactor}{4}
\providecommand{\BIBentryALTinterwordspacing}{\spaceskip=\fontdimen2\font plus
\BIBentryALTinterwordstretchfactor\fontdimen3\font minus
  \fontdimen4\font\relax}
\providecommand{\BIBforeignlanguage}[2]{{%
\expandafter\ifx\csname l@#1\endcsname\relax
\typeout{** WARNING: IEEEtran.bst: No hyphenation pattern has been}%
\typeout{** loaded for the language `#1'. Using the pattern for}%
\typeout{** the default language instead.}%
\else
\language=\csname l@#1\endcsname
\fi
#2}}
\providecommand{\BIBdecl}{\relax}
\BIBdecl

\bibitem{Balanis07}
C.~A. Balanis and P.~I. Ioannides, \emph{Introduction to Smart Antennas}.\hskip
  1em plus 0.5em minus 0.4em\relax Morgan and Claypool Publishers, 2007.

\bibitem{Spilker96}
J.~J. Spilker, \emph{Global Positioning System: Theory and Applications}.\hskip
  1em plus 0.5em minus 0.4em\relax American Institute of Aeronautics and
  Astronautics, 1996.

\bibitem{Zeimetz08}
P.~Zeimetz and H.~Kuhlmann, ``On the accuracy of absolute {GNSS} antenna
  calibration and the concept of a new anechoic chamber,'' in \emph{FIG Working
  Week}, Stockholm, Sweden, 2008.

\bibitem{Kunysz10}
W.~Kunysz, ``Antenna phase center effects and measurements in {GNSS} ranging
  applications,'' in \emph{2010 14th International Symposium on Antenna
  Technology and Applied Electromagnetics \& the American Electromagnetics
  Conference}, Ottawa, Canada, July 2010, pp. 1--4.

\bibitem{Muehldorf70}
E.~Muehldorf, ``The phase center of horn antennas,'' \emph{IEEE Transactions on
  Antennas and Propagation}, vol.~18, no.~6, pp. 753--760, November 1970.

\bibitem{Rong07}
K.~Rong, S.~Donglin, and Y.~Zhengguang, ``The research of phase center of
  log-periodic dipole antenna,'' in \emph{2007 International Symposium on
  Microwave, Antenna, Propagation and EMC Technologies for Wireless
  Communications}, Aug 2007, pp. 735--738.

\bibitem{Betjes07}
P.~N. Betjes, ``An algorithm for automated phase center determination and its
  implementation,'' in \emph{AMTA}, St. Louis, MO, Nov. 2007.

\bibitem{Haerke17}
D.~Harke, H.~Garbe, and P.~Chakravarty, ``A system-independent algorithm for
  phase center determination,'' in \emph{2017 International Symposium on
  Electromagnetic Compatibility - EMC EUROPE}, Sep. 2017, pp. 1--5.

\bibitem{Basta09}
N.~Basta and A.~Dreher, ``On antenna phase centre and its determination with
  nelder-mead simplex optimising algorithm,'' in \emph{2009 European Microwave
  Conference (EuMC)}, Sep. 2009, pp. 854--857.

\bibitem{Almasri17}
S.~A. Almasri, N.~Doose, and P.~A. Hoeher, ``Parametric direction-of-arrival
  estimation for multi-mode antennas,'' in \emph{14th IEEE Workshop on
  Positioning, Navigation and Communications (WPNC '17)}, Bremen, Germany, Oct.
  2017, pp. 1--5.

\bibitem{Poehlmann17WPNC}
R.~P\"ohlmann, S.~Zhang, T.~Jost, and A.~Dammann, ``Power-based
  direction-of-arrival estimation using a single multi-mode antenna,'' in
  \emph{14th IEEE Workshop on Positioning, Navigation and Communications (WPNC
  '17)}, Bremen, Germany, Oct. 2017, pp. 1--6.

\bibitem{Poehlmann17CAMSAP}
R.~P\"ohlmann, S.~Zhang, K.~A. Yinusa, and A.~Dammann, ``Multi-mode antenna
  specific direction-of-arrival estimation schemes,'' in \emph{7th IEEE
  International Workshop on Computational Advances in Multi-Sensor Adaptive
  Processing (CAMSAP '17)}, Curacao, Dutch Antilles, Dec. 2017, pp. 1--5.

\bibitem{Poehlmann19}
R.~{P\"ohlmann}, S.~A. {Almasri}, S.~{Zhang}, T.~{Jost}, A.~{Dammann}, and
  P.~A. {Hoeher}, ``On the potential of multi-mode antennas for
  direction-of-arrival estimation,'' \emph{IEEE Transactions on Antennas and
  Propagation}, vol.~67, no.~5, pp. 3374--3386, May 2019.

\bibitem{Almasri19}
S.~A. {Almasri}, R.~{P\"ohlmann}, N.~{Doose}, P.~A. {Hoeher}, and A.~{Dammann},
  ``Modeling aspects of planar multi-mode antennas for direction-of-arrival
  estimation,'' \emph{IEEE Sensors Journal}, vol.~19, no.~12, pp. 4585--4597,
  June 2019.

\bibitem{ieeestandard}
``{IEEE} standard for definitions of terms for antennas,'' \emph{IEEE Std
  145-2013 (Revision of IEEE Std 145-1993)}, pp. 1--50, March 2014.

\bibitem{Garbacz71}
R.~J. Garbacz and R.~H. Turpin, ``A generalized expansion for radiated and
  scattered fields,'' \emph{IEEE Transactions on Antennas and Propagation},
  vol.~19, no.~3, pp. 348--358, May 1971.

\bibitem{Harrington71}
R.~F. Harrington and J.~R. Mautz, ``Theory of characteristic modes for
  conducting bodies,'' \emph{IEEE Transactions on Antennas and Propagation},
  vol.~19, no.~5, pp. 622--628, Sep. 1971.

\bibitem{Chen15}
Y.~Chen and C.-F. Wang, \emph{Characteristic Modes: Theory and Applications in
  Antenna Engineering}.\hskip 1em plus 0.5em minus 0.4em\relax Wiley, 2015.

\bibitem{martens14}
R.~Martens and D.~Manteuffel, ``Systematic design method of a mobile multiple
  antenna system using the theory of characteristic modes,'' \emph{IET
  Microwaves, Antennas and Propagation}, vol.~8, no.~12, pp. 887--893, Sep.
  2014.

\bibitem{Svantesson02}
T.~Svantesson, ``Correlation and channel capacity of {MIMO} systems employing
  multimode antennas,'' \emph{IEEE Transactions on Vehicular Technology},
  vol.~51, no.~6, pp. 1304--1312, Nov. 2002.

\bibitem{Hoeher17}
P.~A. Hoeher and N.~Doose, ``A massive {MIMO} terminal concept based on
  small-size multi-mode antennas,'' \emph{Transactions on Emerging
  Telecommunications Technologies}, vol.~28, no.~2, Feb. 2017.

\bibitem{Lagarias98}
J.~C. Lagarias, J.~A. Reeds, M.~V.~T. Heckler, and A.~Hornbostel, ``Convergence
  properties of the nelder--mead simplex method in low dimensions,'' \emph{SIAM
  Journal of Optimization}, vol.~9, no.~1, pp. 112--147, 1998.

\bibitem{Manteuffel16}
D.~Manteuffel and R.~Martens, ``Compact multimode multielement antenna for
  indoor {UWB} massive {MIMO},'' \emph{IEEE Transactions on Antennas and
  Propagation}, vol.~64, no.~7, pp. 2689--2697, Jul. 2016.

\end{thebibliography}

\end{document}